*Article*

# Bent Crystal Design and Characterization for High-Energy Physics Experiments

**Marco Romagnoni [1]**, **Vincenzo Guidi [2,1,\*]**, **Laura Bandiera [1]**, **Davide De Salvador[3−4]**, **Andrea Mazzolari[1]**, **Francesco Sgarbossa [3−4]**, **Mattia Soldani[2,1]**, **Alexei Sytov[5,1]** and **Melissa Tamisari[6]**

1    INFN Ferrara Division, Via Saragat 1, 44122 Ferrara, Italy
2    University of Ferrara, Department of Physics and Earth Science, Via Saragat 1, 44122 Ferrara, Italy
3    University of Padua, Department of Physics and Astronomy, via Marzolo 8, 35131 Padova, Italy
4    INFN Legnaro National Laboratories, Viale dell'Università 2, 35020 Legnaro, Italy
5    Korea Institute of Science and Technology Information, 245 Daehak-ro, Yuseong-gu, Daejeon, 34141, Korea
6    Department of Neuroscience and Rehabilitation, Via Luigi Borsari 46, 44121, Ferrara, Italy
*    Correspondence: vincenzo.guidi@unife.it

**Abstract:** Bent crystal are widely used as optics for X-rays, but via the phenomenon of planar channeling they may act as waveguide for relativistic charged particles beam as well, outperforming some of the traditional technologies currently employed. A physical description of the phenomenon and the resulting potential for applications in a particle accelerator is reported. The elastic properties of the anisotropic crystal lattice medium are discussed, introducing different types of curvature which can enable a wide array of bending schemes optimized for each different case features. The technological development of machining strategy and bending solutions useful for the fabrication of crystals suitable in high energy particle manipulations are described. As well as the high precision characterization processes developed in order to satisfy the strict requirements for installation in an accelerator. Finally, the characterization of channeling phenomenon in bent crystal is described, pointing out several experimental setups suitable to comply each specific case constrains.

## 1. Introduction

Single crystals fabricated with extremely high degree of perfection, readily available for example from the semiconductor industry, open a new field of physics research and applications. Indeed, such crystal can be exploited in high energy particle accelerators as a novel type of optics to manipulate charged particles. The phenomenon of planar channeling allows bent crystal to act as a compact and extremely powerful waveguide for deflecting charged relativistic particles without any energy consumption. The design and fabrication of bent crystals suitable for such application requires careful investigation of the prime crystalline material and in-depth knowledge of elastic anisotropy features of the lattice. Dedicated techniques have been developed in order to obtain and characterize the deformational state of the crystals with nanometric precision. Optical methods from laser interferometry usually exploited to characterize mirrors have been adapted, as well as x-rays diffraction techniques used both in laboratories with High-Resolution XRD and in synchrotron light facilities. The characterization of the crystal samples channeling features requires the employment of relativistic particle beam at accelerator facilities and specific beam diagnostic techniques. This work will divulge the progresses accomplished in such fields

## 2. Channeling Phenomenon

The crystal lattice structure was observed for the first time thanks to the effect of X-rays diffraction, a phenomenon impossible in an amorphous medium. Since then, several phenomena exclusively of crystal medium have been theorized and discovered. In particular, the motion of a charged particle travelling in close alignment with an atomic planes (or rows) was found fundamentally different from the general case in an amorphous medium.

This effect is known as channeling and occurs because of the correlation between single scattering events occurring between a charged particle and the screened nuclei placed along atomic planes or strings. In 1964 Lindhard [1] developed the theoretical bases of the phenomenon, introducing the concept of continuous potential to describe the interaction of projectile particles with whole sets of atoms.

For positive particles analytical calculation results in potential wells with maxima on planes/axes and minimum between them (see Fig.1), whereas the opposite takes place for negative charged ones. Depth of such potential wells are of the order of $10^{1-2}eV$, with width of the order of $10^{-10}m$ (typical of interatomic distances in solids), hence introducing very large electric fields $10^{9-10}GeV/cm$.

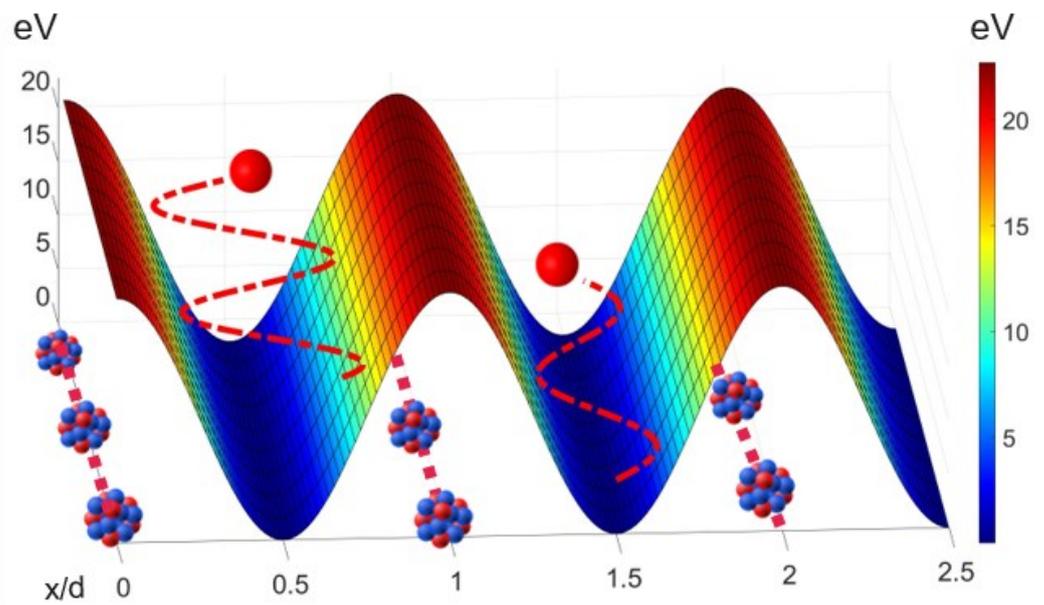

**Figure 1.** Sketch of atomic planes continuous potential and channeling for positive particles, the horizontal axis unit $x/d$ indicate the direction $x$ perpendicular to the lattice planes, normalized by the distance $d$ between two adjacent atomic plane. The y axis indicate the direction parallel to the lattice plane. The potential shows maxima along the nuclei on the atomic planes, letting positively charged particles oscillate between two adjacent planes while crossing the crystal.

Channeling occurs for particle impinging into the crystal, parallel to crystallographic planes and axis within angle lower than the critical one $\theta_c$.

$$\theta_c = \sqrt{\frac{2U_0}{pv}} \tag{1}$$

Where $U_0$ is the potential well depth, $pv$ is the momentum of the particle.

Once in channeling condition, the particle is bounded to the continuous potential of the atomic planes/axes. Main sources of dechanneling events, i.e. particle loosing channeling condition, can be split between incoherent scattering with nuclei and with electrons. The former occurs for a particle travelling near the nuclei, this effect is the most intense as a single scattering event in this case may expel the particle from channeling. The latter is generally caused by multiple scattering events with valence electrons and thus is order of magnitude slower. Resistance of channeling from both effects increase with the momentum of incoming particle.

In particular, the case of positive particle channeling is of great interest as it forces particle to travel in the crystal far from the nuclei in region of low valence electrons density (see Fig.1), thus strongly suppressing the effect of incoherent scattering. Consequently, a channeled beam can cross enhanced length inside a crystal medium while preserving its qualities. Indeed, nuclear dechanneling for positive particles takes place mainly when a particle enters the crystal close to atomic planes and rows.

A method to further suppress nuclear dechanneling for the planar case was devised in [2]. The core concept is to modify channeled particle dynamics by introducing a lattice interruption of precise length at the very beginning of the crystal (either by machining a trench or by burying an amorphous layer [3]). This would allow to exploit the first crystalline layer to focus particles away from nuclei before dechanneling process can take place.(Fig.2)

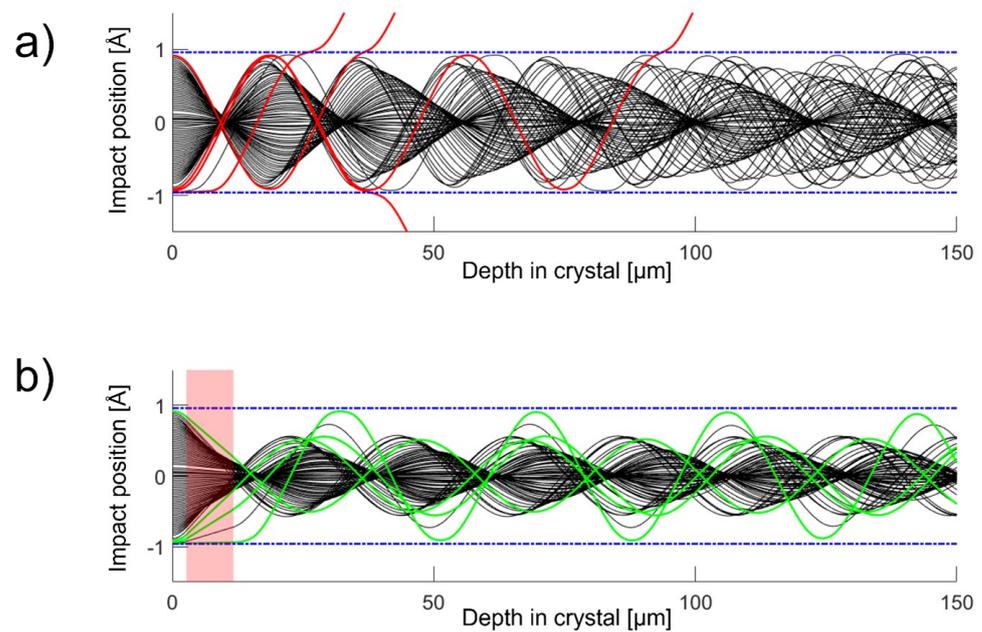

**Figure 2.** MonteCarlo simulation for 180 GeV/c proton planar channeling. Blue lines indicate atomic planes position a) trajectory for normal crystal, red lines indicate dechanneled particles. b) Crystal is interrupted in the pink area, green lines indicate the trajectory of particle subjected to dechanneling in previous case.

An interesting effect can be observed when the crystal presents a curvature along the direction of the channeled particles. In this case, channeled particle will remain bonded to the planes/axes and hence will be forced to follow the crystal curvature (Fig.3). By exploiting this method deflection comparable to hundreds of Tesla magnetic dipole can occur in few millimeters of bent crystal.

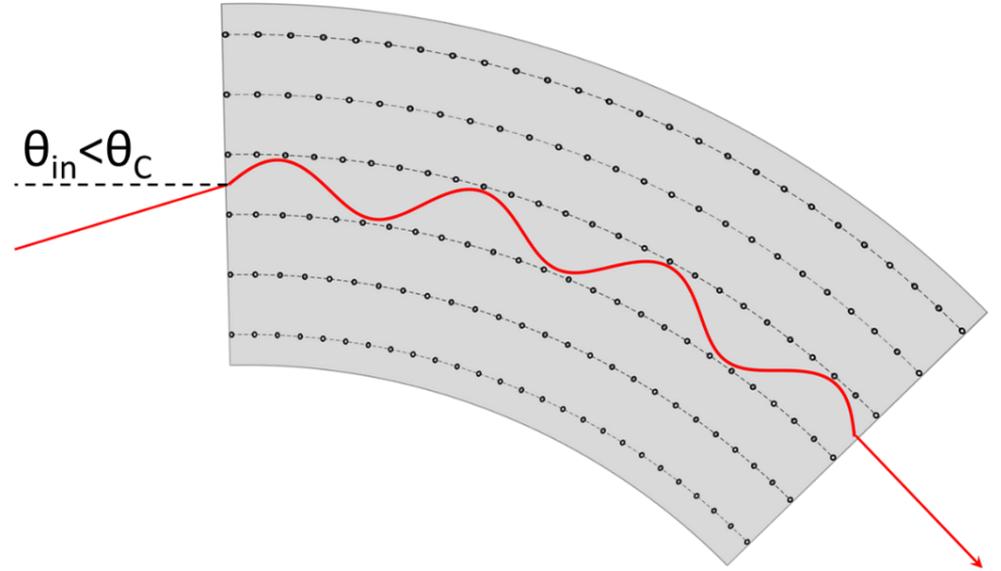

**Figure 3.** Sketch of positive particle trajectory during planar channeling in a bent crystal: if entry angle wrt planes is lower than the critical one $\theta_c$ the particle is channeled between two adjacent atomic planes, following its curvature.

The curvature influence on the continuous potential shape can be described in the reference frame of the particle as an additional centrifugal force.

$$U_{bent}(x) = U_{flat}(x) + \frac{pv}{R}x;\qquad(2)$$

Where $U_{bent}$ and $U_{f\ lat}$ are the continuous potential in bent and flat case, $x$ is the dimension orthogonal to the lattice planes and $R$ is the radius of curvature. This centrifugal component modifies the continuous potential in two ways: it decreases the depth of the potential well and it changes its shapes, introducing an asymmetry. A direct consequence of the first effect is the existence of a limit on the maximum curvature which a channeled particle can follow. For the case of planar channeling, this critical radius can be estimated as

$$R_c = \frac{pv}{U'_{max}};\qquad(3)$$

Where $R_c$ is called critical radius of curvature and $U'_{max}$ is the maximum electric field value in the continuous potential well.

The asymmetry of a bent crystal potential well originates another mechanism for beam steering: the volume reflection. The particle in this case is not trapped into the interplanar channel, but it is instead reflected by the continuous potential in the direction opposite of channeled particles. This occurs for particles which are initially not aligned to the crystal and travels over-barrier in the medium. When such particles reach a point of tangency wrt the planes it is reflected back by the continuous potential by a critic angle. This results in an average deflection of two critical angle in the opposite direction wrt the channeled particles case.

The effect, first theorized in [4] and observed in [5], features an inverse behavior wrt deflection via planar channeling (see Tab.1).

|  | **Channeling** | **Volume Reflection** |
|---|---|---|
| Angular Acceptance | $\theta_{critic}$ | $l/R$ |
| Deflection angle | $l/R$ | $\approx 2\theta_{critic}$ |

**Table 1.** $\theta_{critic}$ is the planar channeling critical angle, $l$ is the crystal length, $R$ is the radius of curvature

### 3. Channeling and defects in real crystals

The precise placement of atoms in an almost infinite array is the fundamental which enables the channeling phenomenon. However, real life crystals are subject to different types of imperfections which partially alter that ideal condition. Each different type has different influence over the particle motion within the interplanar channel, and a correct estimation is critical during design of a bent crystal deflector.

Types of defects can be categorized according to the spatial dimension. Zero dimension or point-like defects consist of single misplaced atoms in the lattice. One-dimensional defects consist in a row of atoms interrupting and displacing the rest of lattice, they can be defined as edge or screw dislocations. Similarly, two dimensional faults are atomic planes interrupting the regular lattice structure, such as stacking faults. Three dimensional defects consist of cluster of point-like defects packed close together to create small amorphous regions.

Another important distinction between lattice defects separates the one already existing in the pristine material and the ones generated as consequence of interaction with particle beams. Dislocation and stacking faults are generated during crystal growth and handling. Point-like defects, and amorphous clusters in a minor part, can be product of irradiation from high energy hadrons on crystal lattice, as extensively investigated by the CERN-RD48 (ROSE) collaboration [6–8].

These types of defects contribution to dechanneling scales with $1/E$ and thus they are the least threatening case for ultrarelativistic particles channeling applications[9]. Consequently, channeling phenomenon features an intrinsic radiation hardness, which is critical for most of the applications in particle accelerators.

For instance at U70 accelerator, a bent crystal was used continuously to extract beam halo for ten consecutive years (1989-1999) [10]. Dedicated studies have been carried out specifically to define channeling radiation hardness. A bent crystal was tested under high intensity rate of averaged $2x10^{13}/s$ 70GeV proton at U70 [11]. In order to define the limit dosage for silicon crystal before affecting channeling efficiency a year-long experiment was conducted at SPS in 1996. The total dose of $5x10^{20}$ 450 GeV proton caused a 30% reduction of channeling efficiency [12]. It is important to mention that radiation hardness during channeling may exceed such limit, as nuclear interaction for positive particles would be significantly suppressed. Indeed, a bent crystal operating continuously for mounts at U70 accumulated an estimated total dose of $10^{20}$ 70 GeV proton and did not suffer of any decrease in channeling efficiency [11].

One and two dimensional defects features two different dechanneling mechanism. An energy-independent cause of dechanneling is the direct impact on a nuclear corridor located on the interplanar channel, occurring because of two-dimensional stacking faults and linear dislocations. Similarly to the particle entry in the crystal medium, the impact parameter of the particle on the nuclear corridor is the only relevant parameter in this interaction [9]. This leakage of particles from channeling state occurs only at the very specific location of the defects, thus the overall influence on the entire beam steering is limited.

The second mechanism is related to the long range effect of dislocations on the lattice shape and consequently on the continuous potential. This is the most relevant source of dechanneling and its effect increases along the particle energy, scaling with $E$. The defect can strain a region of the lattice by a quantity quantifiable by the Burger's vector $b$. The effect on the channeled particle motion is the addiction of a centrifugal force analogous to the lattice plane curvature. In the same way, in a coaxial area around the dislocation with radius $r_d \approx b\theta_c$, this force overcome the potential well barrier and particles lose their channeled state.

For application at world highest energy beam of LHC (7 TeV), dislocations are the primary concern. So much that as the particle beam crosses the crystal, in each section of the crystal normal to the beam propagation the density of dislocations allowed is $1/cm^2$ in order to be able to neglect their contribution to dechanneling [13]. This restricts the available material to silicon and possibly germanium single crystal of the highest grade. Diamond crystal can be produced with extremely high quality, but low size and high price prevents this as an available solution. Other single crystal materials with lower perfection grade would inevitably provide lower steering efficiency.

For instance, planar channeling of 400 GeV/c proton beam from a *LiNbO₃* bent crystal featuring $10^4/cm^2$ dislocation was measured at CERN H8 extracted beamline [14]. The deflection efficiency measured was 3%, significantly lower than the expected value of 68%. In the same experiment, volume reflection was studied as well. In this case, the drop in efficiency was significantly lower (from 94% to ≈ 80%). Indeed, volume reflection occur at very localized location in the sample whereas channeling deflection involves the entire sample length, thus the former process is much more resilient to defects than the latter.

## 4. Planar channeling applications with ultrarelativistic beam

### 4.1. Beam Extraction

The idea of exploiting channeling for ultrarelativistic beam manipulation was first proposed by Tsyganov in 1976 [15]. The first application conceived was the extraction of a circulating beam from an accelerator into an external beamline. The first application conceived was the extraction of a circulating beam from an accelerator into an external beamline. The first experimental attempt was performed at Dubna Synchrophasotron [16], where an 11 mm long crystal operating in parasitic mode diverted an 8.4 GeV proton beam at an angle of 35 mrad with an efficiency of $\approx 10^{-4}$. Later, an intensive experimental campaign was carried out at U-70 (IHEP, Protvino, Moscow), where 50 and 70 GeV proton beams [17,18] were extracted with an efficiency of about 1%.

The low extraction efficiencies were ascribed to sub-optimal geometries of the crystals [19] or to a non-optimized experiment set-up. Although the obtained results showed low efficiency, they demonstrated for the first time the effect of particle steering using channeling phenomenon in bent crystals.

Afterwards, an experimental campaign started in the late 1990s at CERN (proton beams of energies of 14, 120 and 270 GeV [20–22]) and Tevatron (proton beam of energy 900 GeV [23,24]). Extraction efficiencies increased (≈ 10% and ≈ 25% respectively), but still remained at values consistently lower than the ones reached with slow extraction approaches (≈98%). Such low values were ascribed to the presence of an imperfect layer on the crystal surface [21,25,26], unwanted parasitic effects in the deformation of the crystal [20,26], and to an inappropriate choice of the length of the crystal [25,27–29].

Indeed, following simulations demonstrated [25,26] that the length of the crystals used at SPS and Tevatron (30 and 40 mm respectively) was optimized for operations in single-pass mode, not in multi-pass. This mode takes in account the probability of channeling for unchanneled particle after one or more loops in the accelerator ring. As a result, crystals about 5 times shorter would have provided extraction efficiencies about 3 times higher. Simulation models proved to be invaluable tools in the design of crystalline detectors and drove the design and manufacturing of a subsequent generation of crystals [30] and bending schemes [11,31]. This progress resulted in a considerable increase of extraction efficiency, which reached the value of 85% [11].

In 2014 Crysbeam ERC project proposed the employment of a long bent crystal to steer beam halo into an external beamline for fixed target experiments. Afterwards other suggestions were made based on the separation of beam halo from primary beam and its consequent use in a dedicated experiment, but exploiting existing detector inside LHC and thus avoiding extraction to an external line [32,33].

More recently, an investigation has been carried out to devise an extraction system for electron beam of a synchrotron (DESY II Booster Synchrotron) at much lower energy from the examples reported before (6 GeV). Whereas the currently employed scheme delivers a tertiary beam after subsequent interactions with two targets, this scheme would improve beam transfer to extracted beamlines, thanks to the direct extraction of electrons from the accelerator ring.

[34]

### 4.2. Beam Collimation

Beside beam extraction, bent crystals may be exploited in accelerators as a novel and upgraded setup for collimation. Indeed, collimation is a delicate operation for a circular accelerator machine, and its efficiency is fundamental to prevent components damages without reducing beam intensity.

The procedure consists in removing particles which drift away from the nominal beam trajectory and form a potentially dangerous halo around the primary beam. The collimation systems use in series of low-density jaws close to the beam to directly intercept the beam halo and scatter it away, followed by dense material to stop and absorb and contain the scattered particle (Fig.4b).

The introduction of a bent crystal to separate the beam halo has the advantage to offer a much more precise control over the beam halo particles trajectory (instead of randomly scatter them away from the beam) and at the same time reduce the production of secondary particle by decreasing the quantity of material near the beam (Fig.4a). The last is of particular importance for hadronic beam and even more so for ionic beam, because of nuclear fragmentation.

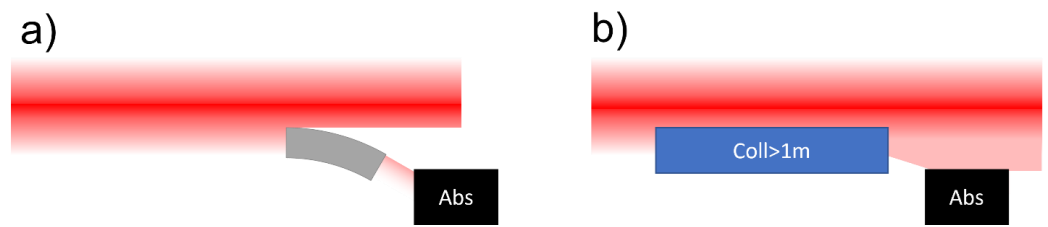

**Figure 4.** a) collimation scheme using a bent crystal, beam halo is precisely directed onto the absorber where it is stopped. b) amorphous collimator scatter beam halo over an angular spread corresponding to the multiple scattering angle to be intercepted by downstream absorber

Crystal assisted collimation for LHC has been the focus of an intense technological development carried out at SPS extracted beamlines. In 2015 the first bent crystal collimator prototypes were installed in the main LHC ring itself, where coherent interactions at record 6.5 TeV energy were observed [35]]. Two more crystal were installed in 2017 and a total of 92h of Machine Development from 2015 until long shutdown 2 in 2018 were dedicated to study and refine crystal use in LHC. Crystal collimation was tested for both protons and ion beam during injection, energy ramp up and at top operation. Positive results led to a first test of crystal collimation not in dedicated test but assisting experiments during collisions. LHC high luminosity upgrade consider employment of crystal for ion beam collimation.

### 4.3. Beam Focusing

Bent crystal can be exploited for more complex beam manipulation, such as focusing of a parallel beam or defocusing from a point like source into a parallel beam.

This can be achieved by modifying the path length (and consequently the deflection angle) with respect to the impact position on the crystal, either by precise shaping (Fig.5a) [36] or by imposing an angle of offset between surface and lattice planes (Fig.5b) [37]. A wide range of focal length can be achieved with these methods, from $10^2 m$ to $10^{-1} m$ [38,38].

For this application bent crystals provide two additional advantages, beside the previously stated large steering power without any energy consumption. One is caused by the small critical angle for high energy particles, which in this case allows to achieve high precision of focus or deliver highly parallel beams. Another advantage wrt quadrupoles is the absence of opposite effects on the direction orthogonal to the focus/defocusing.

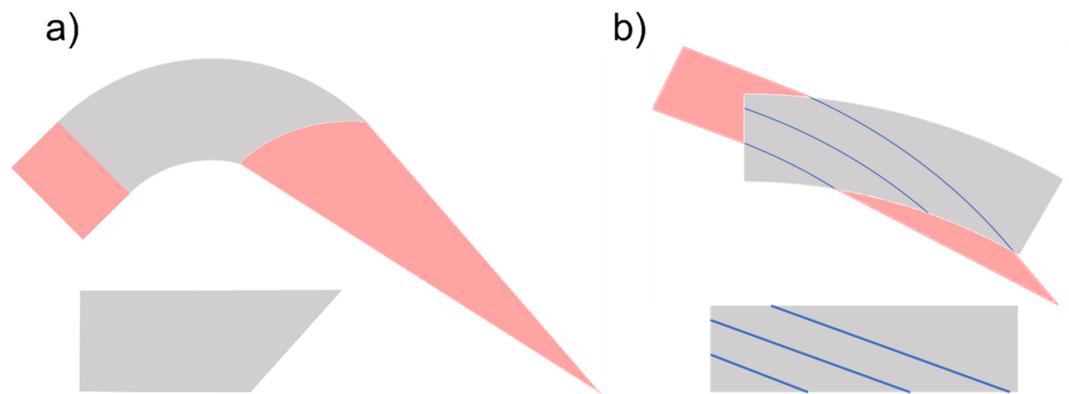

**Figure 5.** a) Focus/defocus crystal, exploiting a machined wedge in order to modify the length of the crystal and consequently the channeling deflection angle in function of the impact parameter. b) In this case, the different path is obtained by a careful selection of miscut angle between sample surface and lattice planes (blue lines)

### 4.4. Spin Precession

During planar channeling, action of the crystal is not limited to modification of trajectory. Indeed, an extremely fast spin precession can occur as well, similarly to the case for a particle in an external magnetic field (Fig.6) [39]. This effect can be exploited to investigate the magnetic dipole moment (MDM) and electric dipole moment (EDM) of a particle. This scheme enables study of fast decaying particles, too short lived to interact a sufficient amount of time in an artificial magnetic field. This was first experimented at Fermilab in 1992 [40], to study hyperion baryon produced in a target by 900 GeV proton beam. More recently, similar studies have been proposed at SPS and LHC for charmed baryon and tau lepton.[33,41]

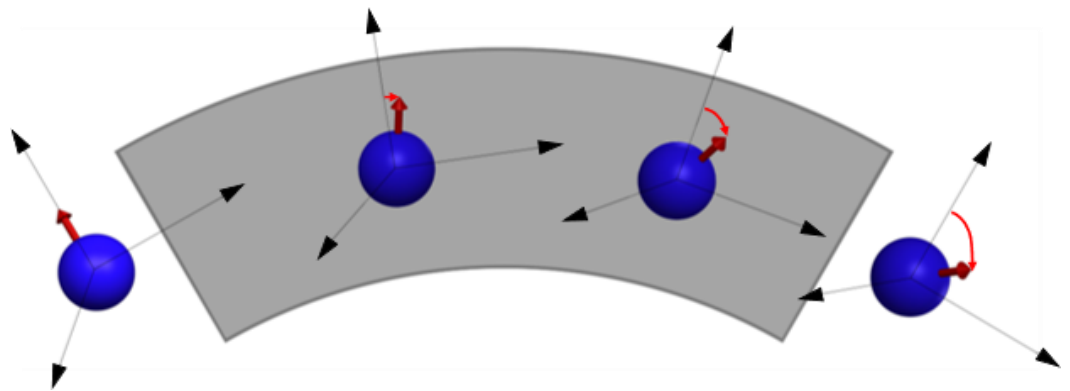

**Figure 6.** Spin precession take place for positive channeled particle in a bant crystal: rotation of spin occurred in the crystal can be exploited to infer magnetic dipole moment or electric dipole moment.

## 5. Crystal Processing

As stated before, in order to ensure maximum channeling efficiency only materials manufactured with perfect lattice quality should be selected. Pristine material can usually be purchased in the form of wafer from electronic provider industry. Silicon single crystal are readily available with very high lattice quality. The starting wafer should be selected with suitable crystallographic orientation and surface features such as roughness, planarity and reference plane deviation.

Crystallographic orientation should be selected in order to exploit the strongest potential. For diamond-like lattice this correspond to (110) planes for positive particles. Unfortunately, quasi-mosaic curvature cannot be achieved along such planes, thus for this configuration the (111) planes are the best choice even if the potential well is weaker.

Secondly, the direction parallel to the channeled particle propagation should be defined. This does not directly influence the planar continuous potential, but significantly affects the mechanical properties of the crystal sample during bending. The ratio between primary and secondary curvatures is directly affected, as well as the presence of undesired effects such as torsion.

In case of beam collimation/extraction, the crystal surface in contact with beam halo should require small roughness. Indeed, beam halo particles slowly drift away from the nominal trajectory they will impact the crystal within few tens' nanometer from the surface. This position, often defined as impact parameter, must be larger than the surface roughness, in order to grant an uninterrupted path of channeled particle along the crystal length. For the same reason, also the surface planarity should be adequately high in order to avoid crossing between surface and lattice planes deeper than the expected impact parameter.

Such features are easily available in top-quality wafers provided by the semiconductor industry. However, even if planarity and roughness requirements are met, another issue is the effect of the reference plane deviation (often referred as miscut) which causes inefficiency in a similar way. Indeed, for small impact parameter lattice planes may interrupt before the end of the crystal for "negative" miscut sign. Moreover, both simulation and experimental evidence during SPS collimation tests indicate a sharp increase of nuclear interaction rate as miscut angle increase. This condition must be avoided in order to safely integrate the use of a bent crystal within an accelerator machine. In particular, for LHC collimation a miscut angle ≤ $10\mu rad$ to minimize nuclear interaction.

This value exceeds of 1-2 order of magnitude the precision achievable by semiconductor industry, thus a dedicated procedure was developed by INFN and University of Ferrara.

First, a refined setup for reference plane deviation measure was devised coupling x-rays diffraction measurement and laser collimator. The wafer mounted on the diffractometer goniometer was aligned to Bragg diffraction before and after a 180° rotation around the normal to its surface. The angular shift of the surface recorded by the autocollimator between the two diffraction positions corresponds to twice the miscut angle. This setup can neglect both mechanical imprecision during rotation and mounting imprecisions, delivering reference plane deviation within $2\mu rad$ resolution.

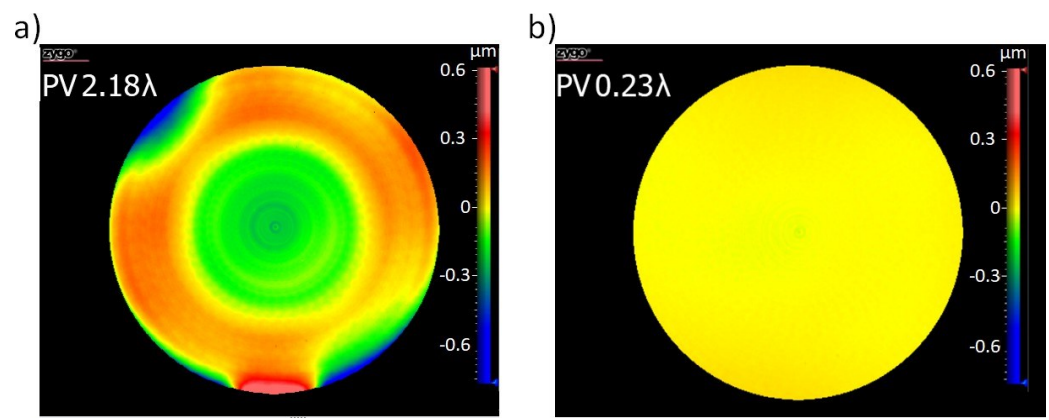

**Figure 7.** Surface processing on wafer surface, interferometric measures of wafer surfacer before (a) and after (b) MRF. Thanks to MRF high planarity can be achieved

To carry out miscut reduction, material must be removed from the surface with extreme precision in order to directly expose the lattice planes. Magnetorheological Finishing (MRF) technology provides the required precision; the technique exploits magnetorheological fluid, whose unique property is that its viscosity changes by several orders of magnitude when introduced into a local magnetic field, essentially turning from a liquid to a quasi-solid in milliseconds. During operation, the fluid is pumped on a rotating spherical wheel which spins it onto the sample surface. An electromagnet beside the spinning wheel adjusts the fluid viscosity and consequently the removal rate. Thus, planarity can be increased by several order of magnitude (Fig.7), as well as alignment of surface with with lattice planes.

| Dislocation Density | Roughness | Planarity (TTV) | Miscut |
|---|---|---|---|
| $< 1/cm^2$ | $100nm$ | $< 0.2 \mu m$ | $< 10 \mu rad$ |

**Table 2.** Requirements on wafer before shaping of crystal specimen for LHC collimation

Once the wafer has been properly prepared with right parameters (see Tab2), the crystal with suitable size and shape for experiments must be obtained. We report techniques devised for this purpose. The crystal shape is tailored according to the goals of the experiment, such as the deflection deflection angle $\theta$. Nevertheless some hard constrains on the sample maximum ($l_{max}$) and minimum ($l_{min}$) length along the beam derive directly from planar channeling physics parameters, such as the critical radius ($R_c$) and the electronic dechanneling length ($L_{dech}$)

$$\begin{cases} l_{min} = \theta R_c \\ l_{max} = L_{dech} \end{cases} \tag{4}$$

Indeed, for curvature beyond the critical radius, channeling becomes impossible. In order to accomplish high channeling efficiency, bending radius should be at least 5 times the critical one. At the same time, most of the particle would escape channeling after crossing one electronic dechanneling length. Consequently, as a general rule of thumb, as particle momentum increases lower curvature and longer crystals becomes more favorable, whereas large curvature and thin crystal becomes more efficient as energy decreases. The following table 3 shows some examples of crystals existing designs.

| Accelerator | Particle Momentum [GeV/c] | Crystal Length [mm] | Deflection Angle [mrad] |
|---|---|---|---|
| LHC | 6500 | 4 | 0.05 |
| FermiLab | 900 | 39 | 0.64 |
| SPS | 400 | 2 | 0.17 |
| U70 | 70 | 2 | 0.9 |
| SLAC | $3 - 4 - 6 - 10 - 14$ | 0.06 | 0.4 |
| MAMI | 0.855 | 0.015 | $0.82 - 1.2 - 1.43$ |

**Table 3.** Example of crystal parameters in past experiments

To avoid damaging the lattice, a completely chemical techniques may be used in some case. Indeed, Silicon micro-machining industry, as a mean to achieve 3D structures, has extensively employed anisotropic etching with different rate along (111), (110) and (100) planes. In particular, potassium hydroxide $KOH$ solutions may feature a ratio of the order of $1 : 100$ between (111) and (110). Exploiting this effect, samples can be etched from the initial wafer with highly planar etched surfaces, provided their parallelism to (111) planes. The procedure requires preliminary steps to prepare the wafer to etching. Through low-pressure chemical vapor deposition (LPCVD), a protective $Si_3N_4$ film $100nm$ thick is deposited on the entire surface of the wafer. With standard photolithographic techniques the desired shape of the samples is masked with photoresist. The protective $Si_3N_4$ film is removed from the exposed areas, leaving them vulnerable to the following etching by $KOH$ solution (20% weight concentration) at $80°C$. Afterwards the remaining $Si_3N_4$ is removed with acid etching.

This technique is also suitable to obtain thin samples from Silicon on insulator (SOI) wafer. The SOI structure is employed in micro-electronic and consist of three layers: a thin film of single crystalline silicon (device layer) is separated from a thick silicon substrate (handle layer) by an electrical insulator, usually silicon dioxide ($SiO_2$). The final sample is obtained from the device layer, which is separated from the handle layer during the last etching step as the $SiO_2$ is removed together with $Si_3N_4$. This type of sample is important for channeling of sub-GeV negatively charged particles, whose dechanneling length is in the order of few tens of microns.

Direct cut of the wafer can be accomplished directly with dicing saws. These instruments provide complete freedom over the crystallographic orientation selection, but inevitably cause a damaged zone along the cuts. Hence, the remotion of such layer is of primary importance for channeling application, especially in case of extremely high-energy particles such as during LHC beam collimation. This process may be carried out by isotropic etching of the sample, taking care to protect the rest of the sample surfaces with deposition of a protective inert layer such as $Si_3N_4$ before the cutting. The damage can be mechanically removed via lapping and polish of the cut surfaces. In this case no film deposition is required prior cutting and low roughness can be achieved.

The quality of the crystal lattice surface throughout the various procedure described, should be kept monitored. In particular, the lattice quality on the superficial layers after MRF and after cut damage removal have been confirmed with several techniques. Each study investigates different volume of material, allowing probing of lattice from deeper to more superficial layers.

X-ray's diffraction of highly collimated and monochromatic radiation is sensible to presence of dislocations and stresses on the lattice. Such defects induce increased signal on the tails of diffraction profiles (rocking curves) obtained during rotation of the sample in and out of diffraction. A comparison of rocking curves measured on processed surfaces and unprocessed material shows the same profile, thus a similar lattice quality integrated over the $\approx 13\mu m$ thickness probed by the employed $8.04 keV$ x-rays photons ($K\alpha_1 Cu$).

The lattice of the processed surfaces produces the same phononic features of unprocessed material as well, as observed by micro-Raman spectroscopy measurements. The analysis of the acquired spectra of optical light ($532nm$) probing the superficial $\approx 1.3\mu m$ perfectly follows the peaks expected for Silicon crystal.

Investigation using Rutherford Backscattering (RBS) of $2MeV$ $^4He^+$ particles over a similar volume (up to $\approx 2\mu m$ depth) confirmed the quality of the lattice after processing. The estimation was performed by analysis of backscattering yields of channeled particles since channeling is sensitive to crystal defects. In particular, the ratio of the yield in case of channeled (i.e. aligned to crystallographic plane/axis) and unchanneled beam (i.e. misaligned to crystallographic plane/axis) was compared for the case of processed and unprocessed surfaces, showing no difference between the two cases.

Finally, the direct observation of the most superficial atomic planes of the crystals by means of high-resolution transmission electron microscopy confirmed as well the presence of a perfect lattice structure uninterrupted until the end of the sample.

## 6. Crystal Bending

A strong point of crystal deflectors is the availability of several bending schemes developed in years of research, which allow to tailor the size of the specimen to the geometrical constrains during operations and experiments.

### 6.1. Primary Curvature

For primary curvature is intended the deformation directly imposed on the crystal. This configuration is useful in order to maximize total deflection and simultaneously minimize the total size of the crystalline deflector, as there are no constraints on the aspect ratio and thus the directions perpendicular to the curvature can be designed as small as needed for beam interception and handling. The nature of the deformation can be of either elastic or plastic origin. Elastic primary curvature may be imposed by a mechanical application of forces to generate a bending momentum using metal bender or by the deposition of a thin tensile film [42] or a pattern of carbon fiber [43] on the surface. Plastic deformation may be applied by generating a thermal gradient or elemental variation of crystal composition [44], or by creating a sacrificial compressed damaged layer by means of grooving [45], lapping [46] and sandblasting [47].

This type of setup has been generally employed on long crystal ($> 1cm$) to impose large deflection angles, usually $> 1mrad$. Such large angles are necessary for applications such as beam extraction.

This was the case for proton beam extraction in several accelerator such as the 8.4 GeV Synchrophasotron [16], 50 GeV U-70 [48], 450 GeV SPS [20] and 900 GeV Tevatron [49]. Primary curvature was taken in consideration as well for 7 TeV proton beam extraction of LHC beam halo, in the Crysbeam project [50]. Another application where primary curvature was successfully exploited is the spin precession of fast decaying particles during planar channeling. At Tevatron the dedicated experiment which proved the feasibility of such phenomenon made use of two bent crystal 4 cm long and bent with multi-point system to deflect hyperion by 1.65 mrad [51].

Currently SELDOM project, which is proposing a conceptually similar experiment at LHC with charmed baryons, selected primary curvature scheme for the bent crystal will be $\approx 12cm$ long and deflect $16mrad$ [52].

Primary curvature is extremely versatile in size and range of bending radius achievable. Selection of crystallographic orientations of the crystal can be freely performed with small influence on final result. The drawback of such approach is the difficulty in achieving a uniform bending radius, as relaxation on the ends of the crystal are challenging to avoid, unless a complete bonding of the entire sample is achieved.

### 6.2. Anticlastic Curvature

The anticlastic curvature is an elastic reaction of the material to a primary curvature, transforming a flat surface into a saddle-shape (see Fig.8). This bending occurs on the same plane of the primary curvature but along perpendicular direction and with opposite convexity. Anticlastic bending occurs is a well-known phenomenon studied for over a century as an unwanted secondary effect during bending of any component, regardless of its crystalline or amorphous nature. In channeling community as well during design of bender for primary curvature a great effort was focused on suppression of anticlastic effects [53].

A configuration of particular interest for the production of crystal for channeling is one which present a relatively small principal curvature imparted by two equal forces at the opposite edges of a parallelepiped sample, whose length is larger than the width and much larger than the thickness. This type of geometry is solved analytically by De Saint-Venant [54] and following studies by Love [55] and Timoshenko [56] using elastic theory on a homogeneous and anisotropic bar under infinitesimal deformations and small displacement. The result indicates the profile $v$ of the bent planes

$$v(x, y, z) = \frac{1}{2S_{33}R}[-S_{13}x^2 - S_{23}y^2 + S_{33}(Lz - z^2) - S_{35}xz];$$ (5)

Where $x$ is the direction along the anticlastic curvature, $z$ is the direction along the primary curvature, $y$ the direction perpendicular to the deformed plane, $L$ the length along the primary curvature, $R$ is the primary curvature radius and $S_{ij}$ are elements of the compliance tensor whose values depends on the crystallographic orientations. An useful free toolkit to implement such calculation is *AniCryDe*, which provide as well estimation on deformation of Si, Ge and diamond plates of adjustable crystallographic orientations and size. [57]

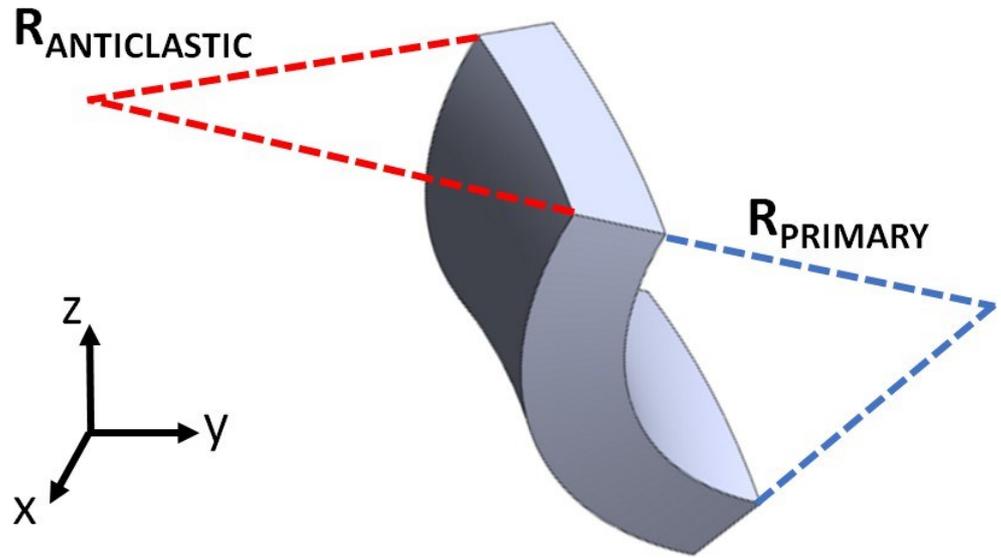

**Figure 8.** sketch of anticlastic bending state

The profile of the anticlastic curvature is a parabola, consequently the anticlastic radius of curvature $R_A$ is constant along the whole length:

$$\frac{1}{R_A} \propto \frac{d^2 v}{dx^2} = -\frac{1}{S_{33}R} S_{13} \tag{6}$$

From this formula is directly obtainable the relation between bending radius of the primary curvature with the anticlastic one. The ratio between the two is called the Poisson ratio, and highlight the strong dependence of the anticlastic curvature on the crystallographic orientation of the sample. Nevertheless, although the bending may significantly change, there is no direction along which the ratio is zero.

$$\frac{R_A}{R} = -\frac{S_{33}}{S_{13}} \tag{7}$$

This ratio has been verified via finite element method and experimentally via curvature measurements with both interferometric and x-rays diffraction on silicon specimen with aspect ratio typical for experiment at CERN. [31] The validity of this equation is constrained by the aspect ratio of the specimen and the primary bending imposed, as indicated by Searle parameter.[58]

$$\beta = \frac{w^2}{Rt} < 1 \tag{8}$$

where $t$ is the thickness of the crystal and $w$ is the width of the crystal along which the anticlastic bending should arise. If $\beta < 1$ anticlastic bending arises on the whole crystal width (so called beam-like behavior); while if $\beta \gg 1$ anticlastic bending remains only in the external regions of the crystal (so called plate-like behavior). In this case, the neutralization of anticlastic bending in a plate is due to the forces that act radially inward toward the center of curvature of a bent sample. As $b$ increases, this radial force becomes large enough to reduce the anticlastic curvature and cause a plane strain condition along the sample width.

Beam-like regime has been widely exploited to produce highly homogeneous curvature on silicon and germanium bent crystal for channeling experiments [58–60]. High channeling efficiency were accomplished for beam extraction and collimation at U70 [11], SPS [61] and LHC [35].

The success of this geometry is based on the possibility to exploit the naturally highly regular central portion of the bent crystal, far from inhomogeneity characteristic of edges (like in case of primary curvature). Another important advantage is the possibility to minimize the quantity of material close to the beam. Indeed, the bent crystal can be easily designed to leave the central portion of the crystal used for channeling very far from any holder structure. This feature can be of critical importance in case of beam collimation, where the least amount of material should be placed near the beam halo in order to avoid perturbation of the main beam dynamic.

As mentioned before, such bending is not suited to produce specimen several cm long. Indeed, to satisfy the Searle parameter condition $\beta \gg 1$, the size of the crystal would scale to massive volume of scarce practical usability and challenging construction requirements. For such cases, borderline conditions to the plate-like behavior should be selected in order to obtain crystals with contained size, as investigated during the R&D of Crysbeam project for possible beam extractor for SPS and LHC [62].

### 6.3. Quasi-Mosaic Curvature

Quasi-mosaic curvature (Fig.9) is a feature exclusive to mechanically anisotropic medium such as crystals. Indeed, it was first observed in quartz crystals during x-rays diffraction [63–65], as a sort of extra mosaicity occurring when stresses were applied on specimen [66], and following studies provided a precise quantitative description of the effect [66]. While for isotropic material is usually valid the Navier-Bernoulli hypothesis which expects during elastic deformation the cross section of the bending planes to remain flat, the anisotropy of the crystal may generate a shear strain which result in a secondary curvature. Only orientation for which compliance tensor of the crystal features mixed elements $S_{ij}$ of high order ($i \vee j > 3$) can produce the shear elastic reaction required for the quasi-mosaic curvature.

A case of particular interest is an anisotropic plate with a couple bending moment $M$ applied on its opposite edges[67]. Defining a reference system with major surface on $xz$ plane, thickness of the plate along the $y$ axis and primary curvature along $z$ axis, the solution for the deformation $u$ along the x-axis is

$$u(x, y, z) = \frac{M}{2I}(S_{12}xy + S_{36}y^2 + S_{35}yz); \qquad (9)$$

Where $I$ is the inertial momentum and $S_{12\text{-}36\text{-}35}$ are elements of the compliance tensor of the material. If $S_{36}$ is not null, quasi-mosaic curvature may occur on $yz$ plane along $y$ direction. The radius of curvature can be directly calculated as

$$\frac{1}{R_{QM}} = \frac{\partial^2 u}{\partial y^2} = \frac{M}{I} S_{36}; \qquad (10)$$

As $S_{36}$ is a function of the lattice orientation, even when the same force is applied the curvature may differ significantly, thus a careful selection of all the three crystallographic orientation is needed in order to define the final curvature. For instance, for $(x, y, z) = (< 2\bar{1}1 >, < 111 >, < 0\bar{1}1 >)$ the 211 planes are bent along $< 111 >$ axis (Fig.9). This configuration would be disadvantageous for planar channeling because of the small potential of 211 planes but enable axial channeling along the strong $< 111 >$ axis. The opposite would occur in a configuration obtained by switching $< 2\bar{1}1 >$ with $< 111 >$. Finally, in case of $x$ axis parallel to $< 0\bar{1}1 >$ no bending can be obtained on $0\bar{1}1$ planes as $S_{36}$ would always be null. This is indeed a drawback of quasi-mosaic bending, as (110) planar potential is the strongest available in diamond like structure like Silicon and Germanium.

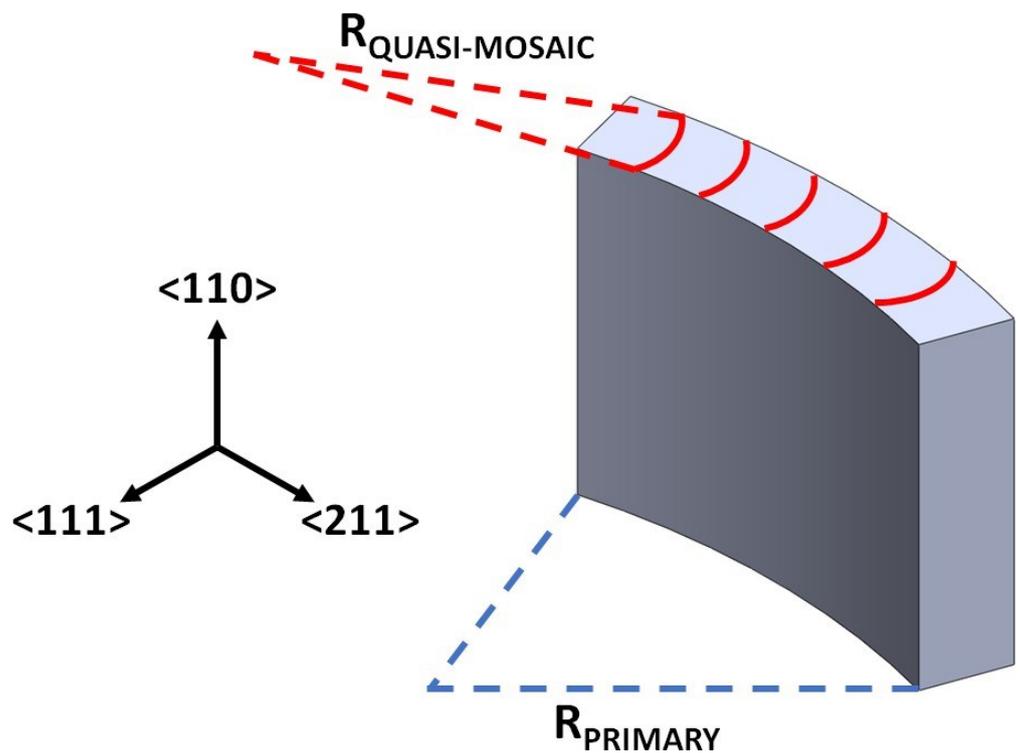

**Figure 9.** sketch of quasi-mosaic bending state

Quasi-mosaic curvature has been introduced to channeling experiments in early 2000s and lead to the first observation of volume reflection at U-70 accelerator [5]. Together with anticlastic curvature is a suitable bending for collimation [35,61].

### 6.4. Final Remarks

The three bending types can be catalogued according to the length of each dimension of the parallelepiped crystal specimen.

Indeed, primary bending would be applied along the longest dimension to provide large deflection angle to extremely energetic particles with large dechanneling length as well as critical bending radius.

Anticlastic bending would be more suitable for the second shorter dimension to perform large deflection of $10^1$ GeV particles or small deflection on more energetic ones ($10^{2-3}$ GeV).

Finally, the quasi-mosaic bending is more naturally imposed on shortest dimension of the crystal. Applications may partially overlap with anticlastic ones, albeit with some important drawbacks. Indeed, the most effective planes for channeling in silicon crystals do not allow quasi-mosaic curvature. Moreover, significant constrain on sample geometry are introduced in order to suppress the effects from anticlastic deformation which would impair channeling angular acceptance. Instead, the most distinctive and effective feature of quasi-mosaic bending is the possibility to attaining large bending along few tens of micrometer lengths. This configuration is required for experiment at "low" energy (few GeV or sub-GeV), especially for negative particles which rapidly escape channeling [68,69].

## 7. Bending techniques

Bending state of the crystal is crucial for channeling performances. Moreover, the working environment often imposes strict requirements over the crystal assembly (size, radiation hardness, vacuum and bake-out compatibility, etc.). Consequently, a careful selection of optimal bending technique is essential during design of any application or experiment. A collection of some techniques used and proposed for channeling experiments is reported.

First experiments exploited multi-points benders, flexing a crystal plate by pressing metal pins on different positions and sides of the sample [70]. This technique is suitable for specimen several cm long, but is not suitable when short crystals and very homogeneous curvature are required (Fig.10).

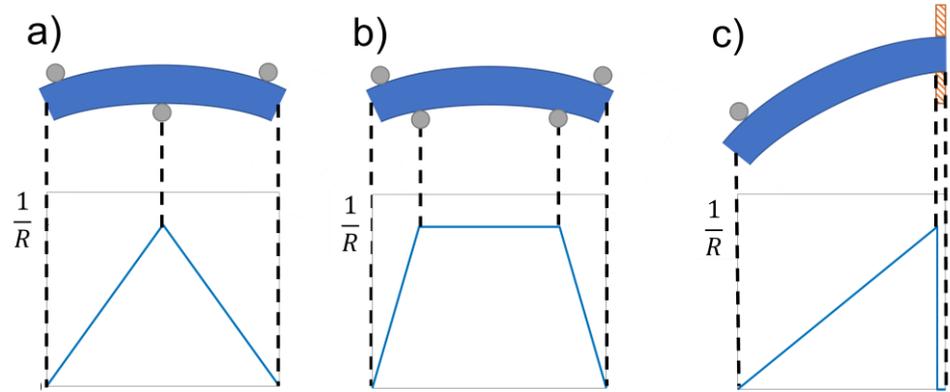

**Figure 10.** Several types of point bender. A crystal slate is bent by using cylindrical pins as fulcrums. A perfectly uniform bending radius cannot be achieved along the entire length of the crystal. [19] .

New types of mechanical holder have been since developed in order to provide more reliable bending. For anticlastic curvature, a series of successful experiments and testing were conducted on SPS extracted lines and for SPS collimation using a metal holder with a mechanism based on screws and bolts. The crystal is clamped at its extremities on two surfaces. The inclination between such surfaces forces the sample to bend, causing indirectly the anticlastic curvature as well. By acting on bolts, the holder is mechanically deformed in order to finely adjust the deformation on the crystal mounted on it. This allows to obtain uniform curvatures as well as correct unwanted elastic reaction such as torsion from crystal anisotropy. The design was first introduced by Russian Institute for High-Energy Physics (IHEP) [11] and further developed by University of Ferrara [71] (see example in Fig.11).

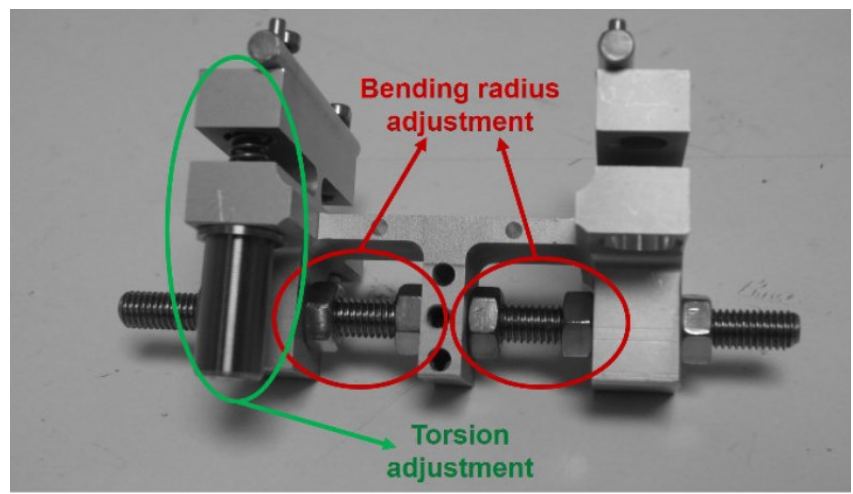

**Figure 11.** A metal holder for anticlastic bending. The crystal sample is clamped on the holder and the system of screws and bolts arch the holder forcing curvature on the sample. A second screw is used to deform the holder in orthogonal direction to the primary curvature in order to correct torsion.

A more recent setup for anticlastic bending was proposed by Ferrara team for collimation in LHC (see Fig.12a-b)[72]. The inclination between the surfaces where the crystal is mounted is machined during holder production, not by following mechanical deformation on the holder. Indeed, the strains and stresses are minimized in order to enhance the stability of the assembly. Such requirement is critical, as the crystal must operate reliably in LHC beampipe without maintenance. In particular, curvature must remain unaffected by the bake-out cycle required to achieve ultra-high vacuum in beampipe (Fig.12c).

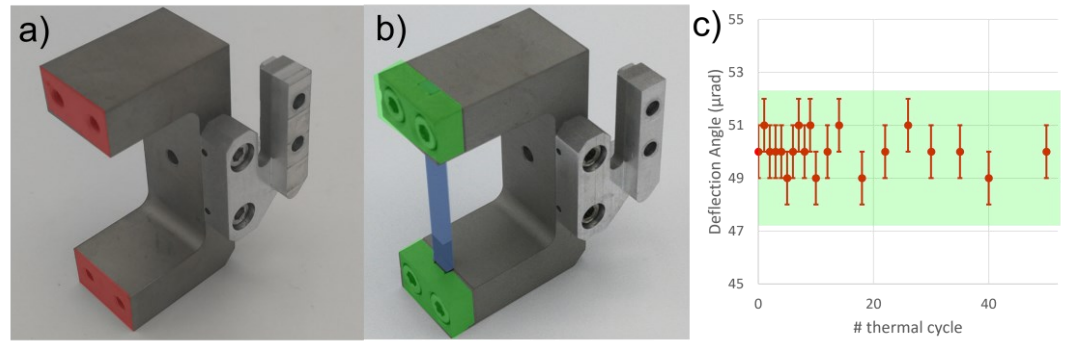

**Figure 12.** a) Picture of holder for LHC collimation, the highlighted red surfaces are inclined. b)A silicon sample (highlighted in blue) is mounted on the red surfaces by clamps (highlighted in green) in order to force an arched shape. c) channeling bending angle is measured after each bake-out process to confirm the assembly stability (light green zone is accepted range for LHC collimation). More details can be found in [72]

Quasi-mosaic curvature is usually achieved clamping a crystal plate between two metal slabs, with profile machined along the desired primary curvature on the sample. The geometry and deformation must be designed in order to suppress as much as possible anticlastic curvature, which would reduce channeling acceptance of the crystal similarly to torsion.

An innovative bending scheme (Fig.13) was recently devised and tested for deflection of sub-GeV electron beams [73]. In this case, the crystal specimen is bent similarly to a bow as two opposite ends are brought closer together. The two sides of the crystal are mounted on two supports which are free to rotate with minimal friction in order to accommodate to the specimen curvature and thus are closely free of stress. This allows to reach radii of curvature close to material breaking limit. The translation inducing the curvature is carried out by a piezo-electric motor, while a second motor adjust the tilt between the two supports to avoid torsion. The two motors can be operated remotely during the experiments to carry out measurements at different radii of curvature.

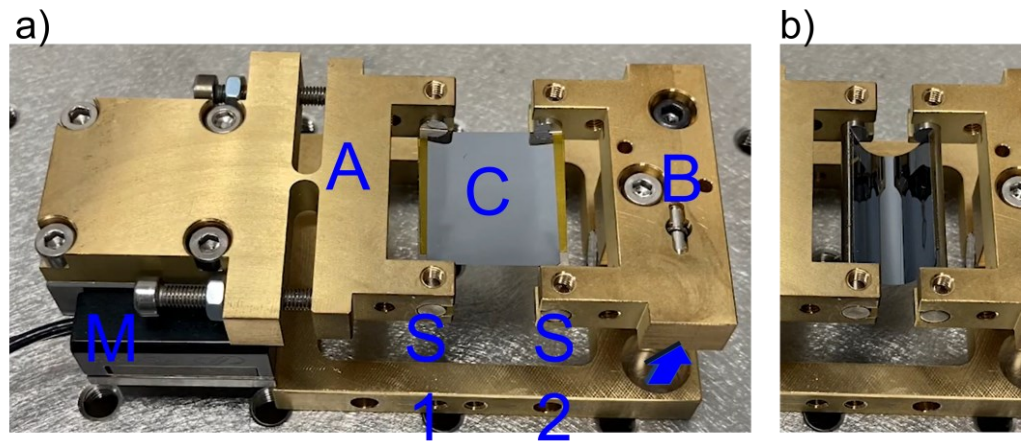

**Figure 13.** a) Picture of the motorized holder for quasi mosaic curvature. Crystal (C) is glued on spines S1-2 and mounted on the holder. The high precision piezo actuated motor (M) moves A towards B, thus forcing the crystal to arch and bend as can be seen in b). A second motor pushing along the arrow can be implemented in order to minimize torsion. Details in [73]

Other techniques discard the employment of a bulky external mechanical holder in order to accomplish a self-standing curvature on the specimen. This can be generally carried out either by deposition of a tensile layer on the sample surface or controlled and localized transformation of the material solid phase.

In the first case, at high temperature a material is deposited on the surface on the specimen. During cooling the different thermal expansion coefficient originate a stress which bend the crystal. The radius of curvature depends on the process parameters (material, temperature and layer thickness) as well as specimen thickness. Thin silicon crystal (few hundreds of microns thick) can be efficiently bent by deposition of few nm of silicon nitride[45], whereas carbon fiber has been proposed for thicker samples (fig.14c)[43].

The use of such techniques depends on the compatibility of the deposited layer with the experimental environment (i.e., vacuum, temperature etc.). The operating temperature, in particular, can affect the stress imposed on the sample and directly affect the curvature.

The second case, i.e., transformation of the specimen material phase, is usually accomplished by a controlled damaging on the surface. Such zones compress the neighbor lattice, inducing a bending stress on the specimen. This process is irreversible, resulting in a permanent curvature on the specimen strongly unaffected by external environment.

Several methods have been proposed such as lapping[46], sandblasting (fig.14b)[47] and grooving (fig.14a)[45]. The main challenge in these procedures is to balance the imposed stress and the damaged area, in order to preserve as much as possible the lattice quality in the rest of the crystal for channeling. Such self-standing methods can achieve remarkably homogeneous curvature and reproducibility, and generally can be easily scaled over a large range of specimen size. Moreover, such techniques are suitable for pattern on the specimen and accomplish complex deformation of the crystal, such as sinusoidal bending for crystal undulators.

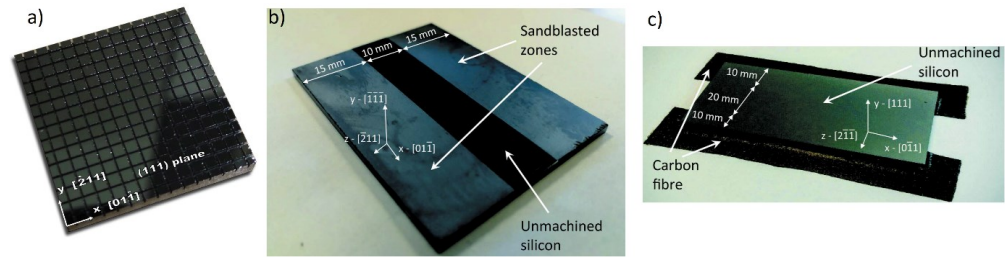

**Figure 14.** Silicon crystals bent using self-standing techniques: a) grooving, b) sandblasting, c) carbon fiber.

This work has focused on techniques developed or exploited for experiments of ultrarelativistic particle channeling. Of course, a plethora of other bending schemes have been invented especially in the field of x-rays optics for synchrotron and more recently for FEL. Very thin diamond and silicon plate (10– $20\mu m$) have been exploited as x-rays spectrometers, using movable mechanics in order to adapt the bending condition. These designs have reached a high level of technological finesse, exploiting carefully designed shapes in order to control sample curvature [74], but still requires nontrivial adjustment in order to be adapted for experiments and application in particle physics. In particular, beam trajectories and deflections are much different between the two cases, as are indeed caused by different phenomena involving different types of particles, thus geometrical encumbrance of crystal assembly should be revised. An important consequence of this is the necessity to use three surfaces in channeling experiments (beam entry face, surface parallel to the beam and beam exit face), whereas only one surface is generally relevant in x-rays application schemes.Another crucial aspect in channeling application is the previously introduced miscut angle, which must be contained with high precision for application in particle accelerator.

## 8. Bending Characterization

Reliable techniques for measurement of crystal bending are necessary for preparation of a channeling experiment. The precision required on the estimation of the deflection angle can reach the microradian in cases such as LHC collimation. Consequently, the techniques must either directly reach angular resolution on lattice planes tilt/curvature or nanometric spatial precision in determination of crystal shape. The former may be achieved via x-rays diffraction, the latter can be achieved via optical interferometry on the surface.

Curvature of lattice planes can be detected as an increased angular acceptance for x-rays diffraction (fig15). Indeed, diffraction in a perfect flat crystal occurs only in a narrow angular range defined as Darwin width, while in a bent crystal such range is increased of the angle subtended by the lattice planes arc bathed by the x-rays beam. Optimal conditions to measure curvature with this method are found at synchrotron beamlines thanks to the intense and monochromatic hard x-rays beam available, featuring photons capable to penetrate the bulk of the crystal and diffract on the very bent planes that will be exploited during channeling [50]. The uniformity of the curvature can be estimated by the intensity of the diffracted beam during rotation of the sample. Indeed, variation of radius of curvature would lead to increase or decrease in the diffraction intensity [75].

This technique allows direct measure of curvature of lattice planes in the bulk of the crystal, such as for the case of quasi-mosaic curvature [76]. Moreover, at synchrotron beamlines is usually possible to focus/collimate the beam to microscopic size, hence achieving high spatial resolution. A drawback of this technique is the limited resolution of the bending caused by the Darwin width, which typically impede resolution below few units of microradians. Another shortcoming occurs for estimation of bending radius uniformity: in case of variation in the medium along the beam path variation in intensity are also related with different absorption. Such can be the case for measure of curvature on long crystals (several centimeters).

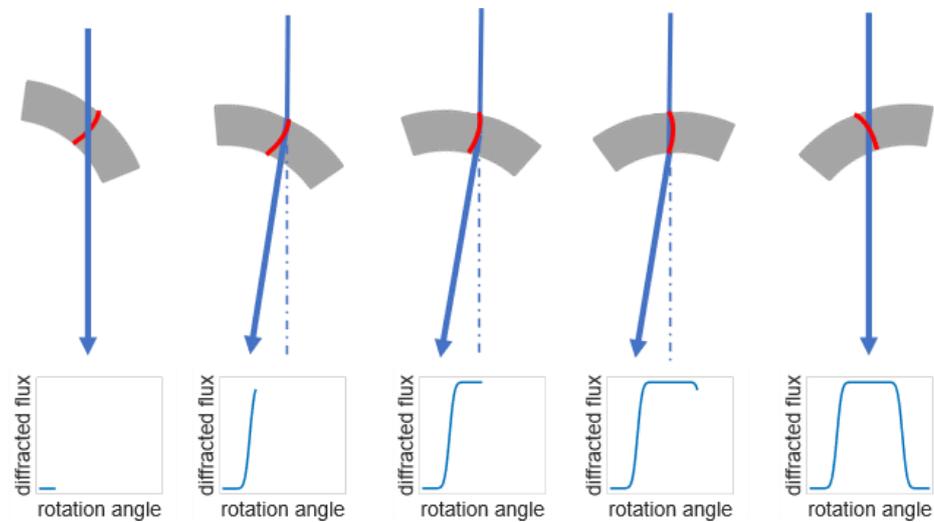

**Figure 15.** Example for Quasi Mosaic curvature (in red). Crystal rotates into and out the diffraction condition, the width of the rocking curve indicates the convolution of Darwin width and lattice planes bending angle.

Whereas the technique described is most suitable for synchrotron facility, x-rays curvature characterization can still be carried out with laboratory instrumentations such as High-Resolution X-Rays Diffractometers (HR-XRD).

In this case a series of rocking curves (RCs) are carried out along the curvature. The angular shift between RCs acquired at different positions is directly caused by the bending of the sample, thus can be used to assess the sample bending radius along the translation direction (see fig.16). This technique requires a reliable rotation and translation system, which is available in Eulerian cradle of high-resolution diffractometers. Another effect to be taken in consideration during such measure is the angular shift caused by pitch and rolling of translation system can introduce artifact in the final results. This type of error can be avoided by the installation of an autocollimator and a flat reference mirror on the sample stage as suggested in [72]. In this case, the autocollimator will measure on the mirror the angular shifts during translation which can be subtracted in the curvature calculations. This allows to reach precision of $1\mu rad$ in the determination of RCs angular shifts, thus allowing the most sensible measures of curvature.

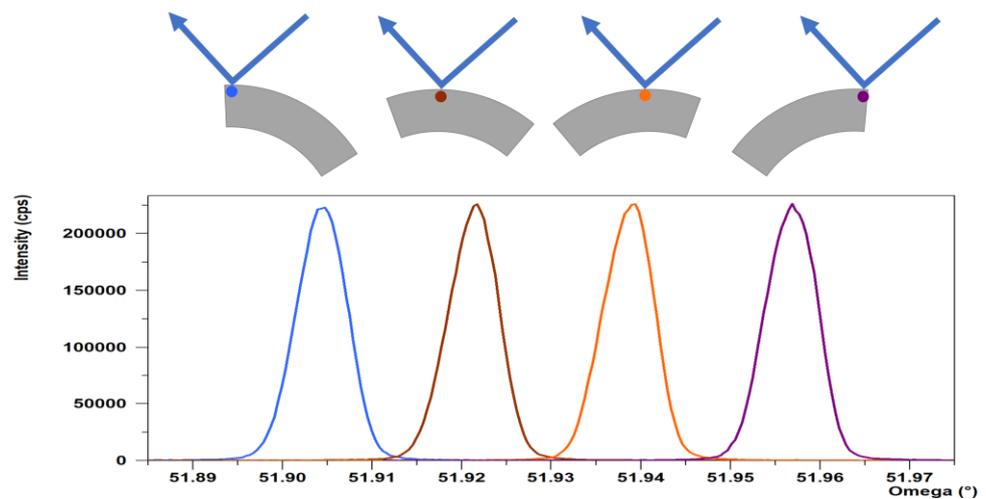

**Figure 16.** RCs are acquired at different positions along the curvature. As lattice planes are bent, tilt of planes changes along the curvature, thus a shift in the centroid of the rocking curves is observed.

Whereas x-rays provide direct measure of the lattice planes, optical characterization of the sample surface can provide useful information. Indeed, as previously stated, the surface of bent crystals for channeling application is often polished to mirror-like state, allowing reflection of incoming light. Optical interferometry is based on the observation of interference pattern obtained between the light reflected by the sample surface and a flat reference ($\lambda/20$) and allows to determine difference in quote on the sample surface with nanometric precision.

This method can be further improved by comparison of surface before and after bending, which allows to exclude the original morphology of the sample surface and thus obtain the pure deformation effect induced by the bending operation. The measure allows to obtain a complete 2D characterization of the bending condition of the entire sample (see fig.17), whereas x-rays diffraction can measure only curvature along one orientation for each characterization. In case of very bent sample (i.e radius < 5*m* ), collection on reflected light back in the interferometer can be partial, thus impeding measure of extended surface. In this case, instead of employing plane wave configuration (i.e. using a flat reference) a more fitting solution would be the use of spherical or cylindrical wave front (by installing lens instead of plane references).

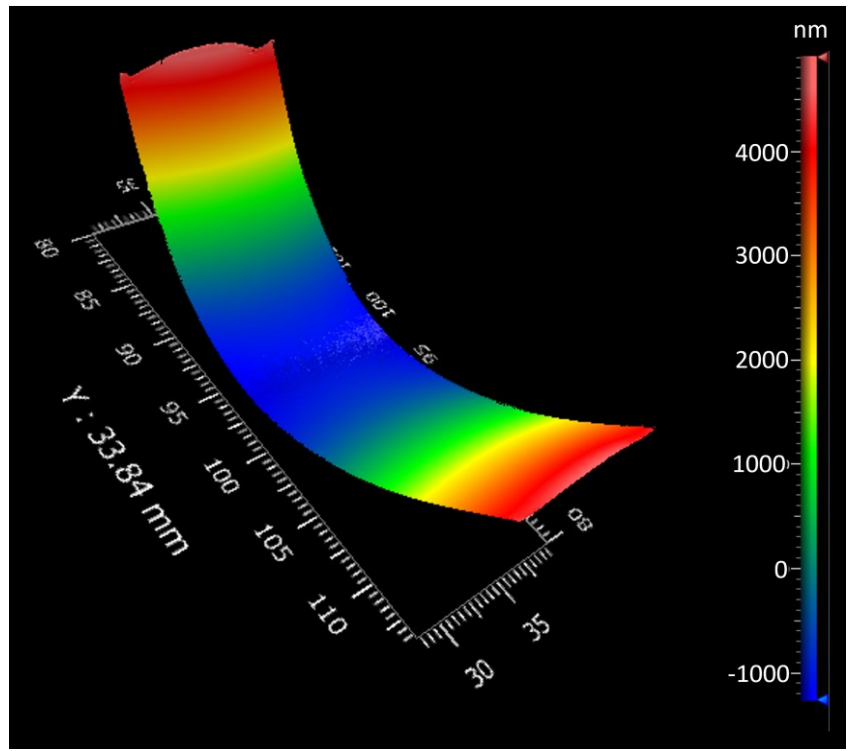

**Figure 17.** Surface profile reconstructed with nanometric precision by laser interferometry

## 9. On-beam Crystal Characterization

Since the first proposal by Tsyganov, experimental characterization of bent crystals performance in high energy particle beam steering have been carried. A direct observation of beam steering by bent crystal can provide information on deflection angle, impact on accelerator machine (i.e. during collimation) and an indication of the efficiency of the channeling effect by defining angular acceptances and fraction of particles successfully channeled along the entire crystal length. This type of analysis is also extremely sensible in the detection of secondary bending effects such as torsion. Moreover, as previously stated, ultrarelativistic particles are an extremely sensitive probe to lattice imperfection in the crystal. As energy increase, both sensitivity and penetration of such particle in the bulk in channeling mode increase, allowing characterization of very thick samples (up to several centimeters). Several different experimental setups have been devised since, adapting and evolving to suite different constrains and exploit new technologies.

The setup can be split into two main functions: an automated handling system to move and align the crystal wrt the incoming particle beams and a detector system to observe the interaction between the beam and the crystal.

The first is constituted by a motorized goniometric system, whose precision must be ideally smaller than the channeling critical angle or at least smaller the beam angular divergence. Nowadays such instrumentation can achieve angular resolution of a microradian, which is suitable for experiments up to TeV energy scale. Beside the required angular resolution, the instrument must also be compatible with the sometime harsh environmental condition. Indeed, accelerators often require operation in vacuum as well as resistance to high dose of radiation.

The beam diagnostic has been in constant evolution, following the technological progress of particle physics detectors in order to achieve the best solutions and adapt to different experimental conditions. First experiments exploited drift chamber to monitor the channeled beam [77], in the 80s and 90s first experimental studies of dechanneling effect exploited energy loss detector built-in on the crystal in order to discriminate which particle started channeling as they enter the crystal [78]. Nowadays the possibility to track each particle of the beam before and after the crystal allows to achieve the highest level of control on the experimental results.

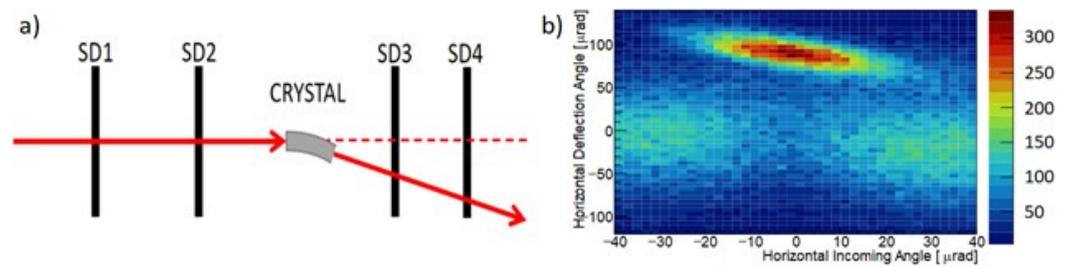

**Figure 18.** a)Scheme of a tracking system to study channeling beam deflection. b) By tracking each particle trajectory before and after the crystal, is possible to study channeling in great detail, such as study deflection in function of the incoming particle angle.

Tracking system (fig.18a) rely on a series of 2d detectors intercepting the beam at different location upstream and downstream the crystal. At each station, the impact position of each particle of the beam on the detector is recorded. Hence, the trajectory of a particle can be inferred as a straight line between two or more consecutive detectors. In case vacuum is maintained between stations, a longer distance allows increased resolution of the trajectory estimation. Instead, if air is present, multiple scattering will progressively deviate stochastically the particle trajectory from a straight line. Thus, an optimal distance should be calculated in order to minimize air contribution and maximize tracker angular resolution. The stations upstream the crystal can be used to measure the incoming angle of the particle into the crystal, while the downstream ones measure the outgoing angle allowing to estimate the beam deflection in function of the incoming angle.

Measure of channeling efficiency can be performed with high precision with this method, allowing for precise compensation of crystal bending imprecision such as torsion or variation of curvature radius wrt to the impact position. Spatial resolution of such detectors is of the order of few tens of microns, thus with sufficiently large distance between detectors is possible to achieve resolution of few units of microradian (fig.18b).

The tracking system solution has been frequently employed in the experiments performed at the extracted beamlines of CERN North Area with very high energy (120 GeV, 180 GeV, 400 GeV), which allowed characterization of crystal before installation in LHC as collimator and first experimental observation of volume reflection [79] and multiple volume reflection in a crystal [80]. The systems used silicon strip detectors to track the particles trajectory, using two detectors upstream the crystal and a variable amount of detector downstream.

Nevertheless this solution is not always neither the most suitable nor feasible approach, depending on the experiment characteristics such as particle type, beam parameters and beamline/accelerator layout. Indeed, this approach is best suited for high energy particles which features negligible energy loss while crossing the detector (i.e. "minimal ionization particle" or "mip"). As energy decreases, the multiple scattering of a particle inside the detector medium increases as well, reducing the angular resolution of the tracking system: if this contribute grows too much wrt the critical angle or the deflection angle, neither a correct estimation of the particle incoming angle and channeling efficiency nor deflection may be possible.

In case a tracking system solution cannot be employed, either for the above-mentioned reasons or because of technical issues of compatibility with the beamline/accelerator layout, channeling experiments have been performed by monitoring only the beam downstream the crystal (example in fig.19), thus losing the information on each particle impact parameters on the crystal (position and angle). For best effectiveness, the detector must be placed sufficiently downstream the crystal in order to allow sufficient spatial separation of the deflected beam from the main one.

Although the drawback of a limited amount of information wrt a complete tracker, this experimental setup has the advantage of being less intrusive on the machine (as it requires the introduction of only one detector) and is well suited for a wide range of energy, as there is no need to preserve beam quality after the detector, so the effect of energy loss and multiple scattering are mostly negligible for the experimental result itself.

This strategy was exploited in study of double channeling in the SPS [81], when 400 GeV proton from the beam halo two was deflected twice by two subsequent bent crystals aligned in channeling in series. A 2d pixeled detector Timepix was placed downstream the second crystal, where it could intercept both the part of the beam channeled once and the beam channeled twice.

Study of channeling efficiency in very short ($15\mu m - 60\mu m$ length) at low energy electrons ($855 MeV$) performed at MAMI X1 beamline monitored the beam downstream the crystal using a scintillator screen [68,82]. The intensity profile resulted in an accurate measure of the beam distribution, although direct absolute count of each single particle was not possible. In this case, the angular divergence of the beam was within the channeling critical angle, allowing precise measurements of channeling efficiency without particle tracking.

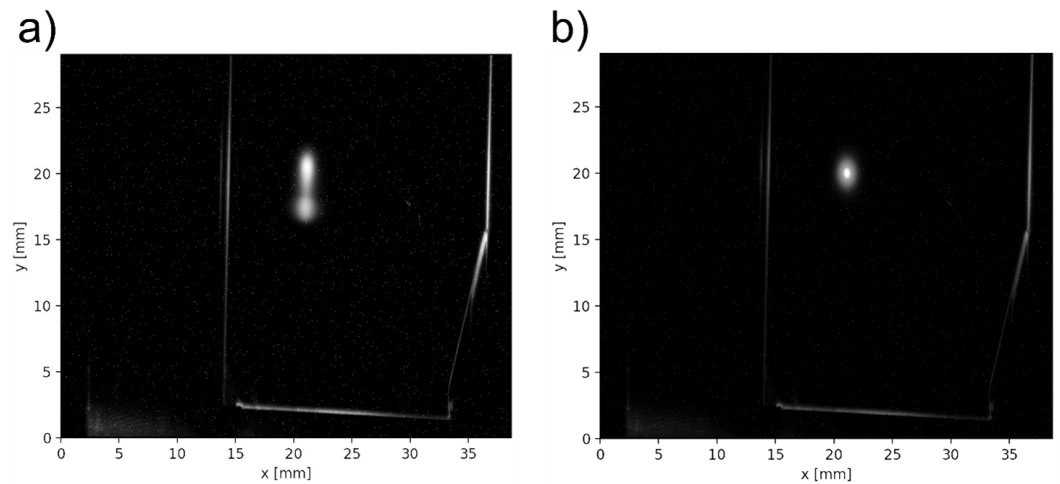

**Figure 19.** Scintillator screen (LYSO) captured by high resolution camera, downstream crystal station. a) channeling is achieved, part of the beam is separated and deflected b) in case the beam is not aligned with crystal, beam profile remains a single spot.

Finally, investigation of bent crystal application in an accelerator can be carried out without a direct interaction with particle of the beam. Such is the case for several studies of beam halo collimation at SPS [61] and LHC [35], which exploited the pre-existing collimators and detectors (beam loss monitor -BLM) installed outside the beampipe for regular diagnostic of beam losses. The alignment of the bent crystal with beam halo and the "scraping" of the channeled particles as a collimator was moved to progressively intercept the steered beam could be detected as variation of signal in BLMs (see fig.20). These measures could not provide a direct measure of the channeling efficiency of the crystal, but could prove nonetheless a clear evidence of channeling effect and could directly monitor the collimation efficiency of the crystal

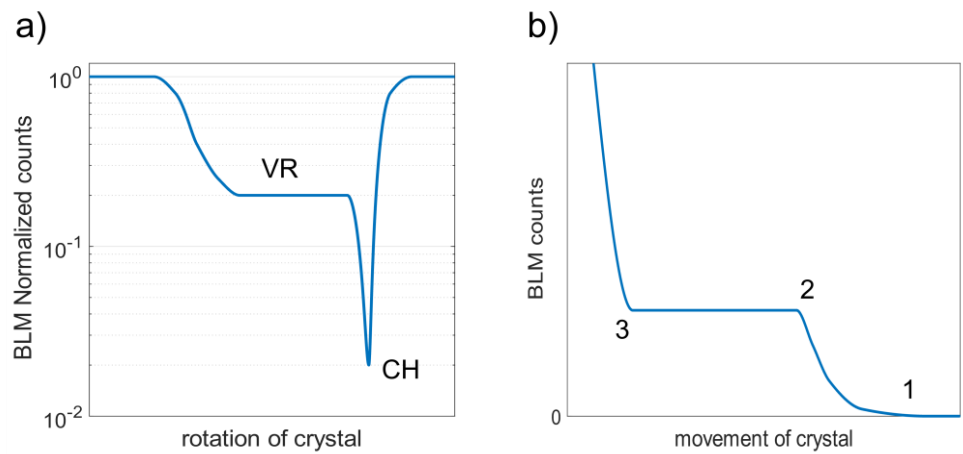

**Figure 20.** Channeling study at LHC by monitoring nuclear interactions of the beam [9] a) signal of the beam loss monitor downstream the crystal during alignment procedure. Decrease in inelastic scattering with crystal is observed during volume reflection (VR) and is finally minimized once in channeling (CH). b) once in channeling, a collimator is moved to gradually towards the beam. Channeled particle are well localized (between 1 and 2) and no more particle are intercepted until close to main beam (3).

## 10. Conclusions

Bent single crystal can be exploited as a powerful and versatile tool to enable novel solutions and schemes in high energy particle accelerators. The development of specialized characterization techniques and the availability of high quality silicon and germanium crystal, allows to obtain perfectly bent crystals with maximum efficiency. Further enhancement of performances may be accomplished only by micromachining structures at the beginning of the crystal to suppress dechanneling effect, as currently investigated by the GALORE project. Other important developments could be enabled if more materials with the required high crystalline quality were to be obtained in the future. Indeed, lattice featuring high atomic number atoms would provide much higher potential and consequently increase the steering power via planar channeling. Piezoelectric material such as *LiNbO*$_3$ could allow a dynamic control of the crystal in order to finely tune its curvature properties, increasing reliability and versatility of bent crystals applications.

**Acknowledgments:** We acknowledge the support from CSN1 and CSN5 of INFN, the INFN project STORM. Partial support by the EU Commission H2020 N-LIGHT e H2020-MSCA-IF TRILLION (GA. 101032975) projects.

**Funding:** This work has been financially supported by the Istituto Nazionale di Fisica Nucleare (INFN), Italy, Commissione Scientifica Nazionale 5, Ricerca Tecnologica Bando No. 23246/2021, project GALORE.

## Abbreviations

The following abbreviations are used in this manuscript:

| | |
|---|---|
| $\theta$ | Channeling deflection angle |
| $\theta_c$ | Channeling critical angle |
| $\theta_{in}$ | incoming angle between particle and lattice plane |
| $U, U_0$ | Continuous potential, Continuous potential depth |
| $U'$ | Electric field value in the continuous potential |
| $R, R_c$ | Radius of curvature, Critical radius of curvature |
| $R_A, R_{QM}$ | Anticlastic and Quasi-mosaic radius of curvature |
| $L_{dech}$ | Electronic dechanneling lenght |
| pv | Particle momentum |
| $b$ | Burger's vector |
| $S, S_{ij}$ | Compliance tensor, Compliance tensor element |
| $\beta$ | Searle parameter |
| MDM | Magnetic Dipole Moment |
| EDM | Electric Dipole Moment |
| LPCVD | Low Pressure Chemical Vapor Deposition |
| SOI | Silicon on Insulator |
| RC | Rocking Curve |
| XRD/HR-XRD | X-Rays Diffractometer/High-Resolution X-rays Diffractometer |
| RBS | Rutherford Backscattering |
| mip | minimal ionization particle |
| LHC | Large Hadron Collider |
| SPS | Super Proton Synchrotron |
| BLM | Beam Loss Monitor |